\documentclass[aps,preprint]{revtex4-1} 

\usepackage{amsmath,amssymb}    
\usepackage[dvipdfmx]{graphicx}   
\usepackage{verbatim}   
\usepackage{color}      
\usepackage{subfigure}  
\usepackage{hyperref}   
\usepackage{here}       
\usepackage{bm}         
\usepackage{color}

\begin{document}

\preprint{KOBE-COSMO-17-15}

\title{Anisotropic Constant-roll Inflation}
\author{Asuka Ito}
\author{Jiro Soda} 

\affiliation{Department of Physics, Kobe University, Kobe 657-8501, Japan}


\begin{abstract}
\noindent \hrulefill 
\begin{center} \normalsize{Abstract} \end{center}

We study constant-roll inflation in the presence of a gauge field coupled to an inflaton. 
By imposing the constant anisotropy condition, we find new  exact anisotropic constant-roll inflationary solutions
which include anisotropic power-law inflation as a special case. 
We also numerically show that the new anisotropic solutions are attractors in the phase space.

\noindent \hrulefill 
\end{abstract}

\maketitle


\section{\normalsize{I\lowercase{ntroduction}}}
It is widely believed that the accelerated expansion of the universe 
in the early universe explains cosmological observations such as the large scale structure of the universe and 
the temperature anisotropy of the cosmic microwave background radiation.
Such a rapid expansion is called inflation~\cite{Starobinsky:1980te}-\cite{Linde:1981mu};
it is driven by an inflaton field $\phi$.
Conventionally, we impose slow roll conditions, namely 
$-\frac{\dot{H}}{H^{2}} , \frac{\ddot{\phi}}{H \dot{\phi}} \ll 1$,  where $H$ is the Hubble parameter 
and an overdot denotes a derivative with respect to the cosmic time.
This is required to realize a sufficiently long ($\sim 60$ e-foldings) 
quasi de Sitter phase and to make the predictions consistent with the observational data.

Recently, constant-roll inflation characterized by a constant-roll condition, 
$\frac{\ddot{\phi}}{H \dot{\phi}} = \rm{const.}$, instead of the slow roll conditions 
has been discussed. 
Under the constant-roll condition, one can solve equations exactly
\cite{Motohashi:2014ppa}.
Thus, one can systematically find a certain class of constant-roll inflationary models which are compatible with 
the observational data \cite{Motohashi:2014ppa}-\cite{Anguelova:2017djf}.
Moreover, one can extend an idea of constant-roll inflation to  teleparallel $f(T)$ gravity where 
the inflaton is coupled to the $f(T)$ \cite{Awad:2017ign}. 
The generalization of the constant-roll condition to  $f(R)$ gravity has been presented
\cite{Nojiri:2017qvx}-\cite{Odintsov:2017hbk}.
Since  $f(R)$ gravity models can be related to  scalar-tensor theory, 
scalar fields play the central role in all these models.

In this paper, we study constant-roll inflationary models in the presence of a gauge field coupled to an inflaton. 
Now, we can expect anisotropy of the expansion of the universe.
Indeed, anisotropically expanding inflationary models have been studied
in the context of the supergravity; they are called anisotropic inflation \cite{Watanabe:2009ct}.
There are various extensions of anisotropic inflationary models \cite{Soda:2012zm}-\cite{Do:2017qyd}
and various predictions for the observations \cite{Gumrukcuoglu:2010yc}-\cite{Bartolo:2017sbu}.
Here, we provide a new way to investigate anisotropic inflation.
Remarkably, 
we find exact  anisotropic constant-roll inflationary solutions which would be attractors in the phase space.

In section II, we present inflationary models we consider in the present paper. In section III, 
we study the constant-roll condition in  anisotropic inflation.
In section IV, we show specific examples to illustrate how to solve the system.
In section V, we investigate the phase space dynamics to
 show that anisotropic solutions are attractors in the phase space.
The final section is devoted to a summary.

\section{\normalsize{I\lowercase{nflation with a gauge field}}}
We consider an inflationary universe with a gauge field. 
The action is given by 
\begin{equation}
    S=\int d^{4}x \sqrt{-g}\left[ 
                     \frac{M_{pl}^{2}}{2}R-\frac{1}{2}(\partial_{\mu}\phi)(\partial^{\mu}\phi)
                          - V(\phi) -\frac{1}{4}  f^{2}(\phi) F_{\mu\nu}F^{\mu\nu}
                              \right] \ ,  \label{action}
\end{equation}
where $g$ is the determinant of the metric $g_{\mu\nu}$ , $M_{pl}$ is the reduced Planck mass, 
$R$ is the Ricci scalar and the inflaton field is represented by $\phi$ with a potential $V(\phi)$.
The field strength of the gauge field $F_{\mu\nu}$ is defined by a gauge potential $A_{\mu}$ as 
$F_{\mu\nu}=\partial_{\mu}A_{\nu}-\partial_{\nu}A_{\mu}$ and it couples to the inflaton field through 
a coupling function $f(\phi)$.

For the gauge field, we take an homogeneous ansatz $A_{\mu}=(0,v_{A}(t),0,0)$ where the specific direction is along with 
the $x$-direction.
In the presence of the non-trivial gauge field, we can not take an isotropic metric. Instead,  we
take the anisotropic ansatz 
\begin{equation}
    ds^{2}=-dt^{2}+a(t)^{2} \Big[\  b(t)^{-4} dx^{2} + b(t)^{2}(dy^{2}+dz^{2}) \  \Big] \ , 
\label{eq2}
\end{equation}
where $a(t)$ describes the average expansion of the universe and $b(t)$ represents the anisotropic expansion of 
the universe.
Now, we can derive equations of motion. It is easy to solve the equation for the gauge field as
\begin{eqnarray}
    \dot{v}_{A}=p_{A}f^{-2} a^{-1} b^{-4} \ .  \label{eq3}
   \label{integral}
\end{eqnarray}
where $p_{_{A}}$ is an integration constant. 
With Eq.(\ref{eq3}), one can obtain the Hamiltonian constraint
\begin{equation}
    H_{a}^{2} - H_{b}^{2} = 
     \frac{1}{3M_{pl}^{2}} \Big[\  \frac{1}{2}\dot{\phi}^{2} + V
                                 + \frac{p_{A}^{2}}{2} f^{-2} a^{-4}b^{-4}  \ \Big] \ ,
                              \label{eq4}
\end{equation}
where $H_{a} \equiv \frac{\dot{a}}{a}\ , \ \ H_{b} \equiv \frac{\dot{b}}{b} $ .
The rest of the Einstein equations are
\begin{eqnarray}
  \dot{H}_{a} &=& -3 H_{b}^{2} -\frac{1}{2M_{pl}^{2}} \dot{\phi}^{2} 
                        -  \frac{p_{A}^{2}}{3M_{pl}^{2}} f^{-2} a^{-4}b^{-4} \ , \label{eq5}  \\
  \dot{H}_{b} &=& -3 H_{a} H_{b} + \frac{p_{A}^{2}}{3M_{pl}^{2}} f^{-2} a^{-4}b^{-4} \ .    \label{eq6}
\end{eqnarray}
The field equation for the inflaton is
\begin{equation}
  \ddot{\phi} = -3 H_{a} \dot{\phi} - V_{,\phi} + p_{A}^{2} f^{-3} f_{,\phi} a^{-4}b^{-4} \ . 
                  \label{klein}
\end{equation}

\section{\normalsize{A\lowercase{nisotropic constant-roll inflation}}}

Now, we impose the constant-roll condition on Eq.(\ref{klein})
\begin{equation}
  \ddot{\phi} = -(3 + \alpha) H_{a} \dot{\phi} \ ,   \label{eq8}
\end{equation}
which is parameterized by a constant $\alpha$.
On top of the constant-roll condition, it is reasonable to seek inflationary solutions with the constant anisotropy condition
\begin{equation}
  \frac{H_{b}}{H_{a}} = n \ ,  \label{eq9}
\end{equation}
where $n$ is a constant. 
From Eqs.(\ref{eq5}),(\ref{eq6}) and (\ref{eq9}), we obtain
\begin{equation}
  (1+n) \dot{H}_{a} = -3n (1+n) H_{a}^{2} - \frac{\dot{\phi}^{2}}{2M_{pl}^{2}} \ . \label{eq10}
\end{equation}
Note that this reduces to the conventional one in the isotropic case $n=0$.
Since one can regard $H_{a}$ as a function of $\phi$, Eq.(\ref{eq10}) can be rewritten as
\begin{equation}
   (1+n) \dot{\phi} \frac{d {H}_{a}}{d\phi} = -3n (1+n) H_{a}^{2} - \frac{\dot{\phi}^{2}}{2M_{pl}^{2}} \ .
\end{equation}
This can be solved with respect to $\dot{\phi}$ as
\begin{equation}
  \dot{\phi} = - M_{pl}^{2} (1+n) \frac{dH_{a}}{d\phi} 
                \pm \sqrt{\Big( M_{pl}^{2} (1+n) \frac{dH_{a}}{d\phi} \Big)^{2} 
                           -6M_{pl}^{2} n(1+n) H_{a}^{2}} \ . \label{eq12}
\end{equation}
Differentiating it with respect to $t$ and using (\ref{eq8}), we get
\begin{equation}
  -(3+\alpha) H_{a} = -M_{pl}^{2} (1+n) \frac{d^{2}H_{a}}{d \phi^{2}} \pm 
          \frac{M_{pl}\sqrt{1+n} \Big( M_{pl}^{2} (1+n) \frac{dH_{a}}{d\phi} \frac{d^{2}H_{a}}{d\phi^{2}} 
                 - 6nH_{a} \frac{dH_{a}}{d\phi}\Big)} 
                   {\sqrt{M_{pl}^{2}(1+n) (\frac{dH_{a}}{d\phi})^{2} - 6nH_{a}^{2}}} \ . \label{eq13}
\end{equation}

Here, let us recall the case of an isotropic universe ($n=0$). In this case, Eq.(\ref{eq13}) is reduced to 
\begin{equation}
  \frac{d^{2}H_{a}}{d\phi^{2}} = \frac{3+\alpha}{2M_{pl}^{2}} H_{a} \ .  \label{eq14}
\end{equation}
%
One can immediately solve Eq.(\ref{eq14}) as
\begin{equation}
  H_{a}(\phi) = C_{1} \exp \Bigg( \sqrt{\frac{3+\alpha}{2}} \frac{\phi}{M_{pl}}\Bigg) +
                C_{2} \exp \Bigg( -\sqrt{\frac{3+\alpha}{2}} \frac{\phi}{M_{pl}}\Bigg)  \ . \label{eq15}
\end{equation}
This is the general solution of isotropic constant-roll (inflation) \cite{Motohashi:2014ppa}.

Given this hint  (\ref{eq15}), we can seek anisotropic constant-roll solutions with the following ansatz:
\begin{equation}
  H_{a}(\phi) = C_{1} \exp \Bigg( \lambda(n)\sqrt{\frac{3+\alpha}{2}}\frac{\phi}{M_{pl}} \Bigg) +
                C_{2} \exp \Bigg( -\lambda(n)\sqrt{\frac{3+\alpha}{2}}\frac{\phi}{M_{pl}} \Bigg)  \ . 
                \label{eq16}
\end{equation}
Substituting this ansatz (\ref{eq16}) into Eq.(\ref{eq13}), we obtain the following algebratic equations:
\begin{eqnarray}
  \lambda = \sqrt{\frac{3+\alpha}{(1+n)(-3n+3+\alpha)}}\quad , \quad  C_{1}C_{2}n(3+\alpha)(3+\alpha-6n) =0 \ .
    \label{eq17}
\end{eqnarray}
%
%
%

In the case of $C_{1}=0$ or $C_{2}=0$, they correspond to 
anisotropic power-law inflation \cite{Kanno:2010nr,Ito:2015sxj}. 
Moreover, $n=0$ means isotropic constant-roll inflation and $\alpha = -3$ represents the de Sitter universe. 
Thus, interesting and new solutions correspond to  the case  $n=\frac{3+\alpha}{6}$.
This tells us that we can reach the de Sitter limit ($\alpha = -3$) by taking an isotropic limit $n \rightarrow 0$. 
It is consistent with Wald's cosmic no-hair theorem \cite{Wald:1983ky},
roughly speaking, which claims an initially homogeneous and anisotropic spacetime approaches the de Sitter spacetime
in the presence of a positive cosmological constant.
It implies that no fields at all can survive and the spacetime becomes isotropic eventually.
In contrast, anisotropic inflation can be realized in slow-roll inflation because it 
deviates from an exact de Sitter spacetime.
The relation $n=\frac{3+\alpha}{6}$ means that anisotropy is related to the deviation from a de Sitter universe, 
this is a feature of anisotropic inflation~\cite{Watanabe:2009ct}.

Hereafter in this paper, we focus on the non-trivial case 
\begin{eqnarray} 
 n=\frac{3+\alpha}{6} \ , \label{consistency}
\end{eqnarray}
which determines the degree of the anisotropy of the spacetime.
In this case, Eq.(\ref{eq17}) yields
\begin{eqnarray}
\lambda = \frac{2\sqrt{3}}{\sqrt{9+\alpha}}
\end{eqnarray}
and Eq.(\ref{eq16}) becomes
\begin{equation}
  H_{a}(\phi) = C_{1} \exp \Bigg( \sqrt{6} \sqrt{\frac{3+\alpha}{9+\alpha}} \frac{\phi}{M_{pl}} \Bigg) +
                C_{2} \exp \Bigg( - \sqrt{6} \sqrt{\frac{3+\alpha}{9+\alpha}} \frac{\phi}{M_{pl}} \Bigg)  \ . 
                \label{eq19}
\end{equation}
Since we have obtained $H$ as a function of $\phi$, the remaining equations are completely integrable.

\section{\normalsize{S\lowercase{pecific examples}}}

Let us illustrate how to solve the remaining equations. 
We consider a special cases of (\ref{eq19}) as
$C_{1}=C_{2}= M/2$, then
\begin{eqnarray}
  H_{a}(\phi) = M \cosh \Bigg( \sqrt{6} \sqrt{\frac{3+\alpha}{9+\alpha}} \frac{\phi}{M_{pl}} \Bigg) \ .
    \label{eq20}
\end{eqnarray}
From Eq.(\ref{eq12}), one can express $\dot{\phi}$ by $\phi$ 
\begin{eqnarray}
  \dot{\phi} = M M_{pl} \sqrt{\frac{-(3+\alpha)(9+\alpha)}{6}} 
       \bigg[ 
       1 + \sin \Big( \sqrt{6} \sqrt{\frac{-(3+\alpha)}{9+\alpha}}\frac{\phi}{M_{pl}} \Big) \bigg] \ .
          \label{eq_phi_dot}
\end{eqnarray}
By solving Eq.(\ref{eq_phi_dot}), we can express $\phi$ as a function of $t$.
Then 
the scale factor $a(\phi)$ is immediately found from the definition $H=\frac{\dot{a}}{a}$ as
\begin{eqnarray}
  a  \propto  \bigg[   
           1 + \sin \Big( \sqrt{6} \sqrt{\frac{-(3+\alpha)}{9+\alpha}}\frac{\phi}{M_{pl}} \Big)
              \bigg]^{\frac{1}{-(3+\alpha)}} \ .
\end{eqnarray}
The potential $V(\phi)$ and the coupling function $f(\phi)$  
are obtained from Eqs.(\ref{eq4}) and (\ref{eq6}):
\begin{eqnarray}
V &=& \frac{M^{2} M_{pl}^{2}(\alpha-9)}{24}  \bigg[ - (6+\alpha) +
             \alpha \cos \Big( 2\sqrt{6} \sqrt{\frac{-(3+\alpha)}{9+\alpha}}\frac{\phi}{M_{pl}}\Big) \nonumber \\ 
             && \hspace{4cm} 
            +  2(3+\alpha) \sin \Big( \sqrt{6} \sqrt{\frac{-(3+\alpha)}{9+\alpha}}\frac{\phi}{M_{pl}} \Big) 
              \bigg] 
\end{eqnarray}
and
\begin{eqnarray}
f \propto  \frac{
           \bigg[ 
           1 + \sin \Big( \sqrt{6} \sqrt{\frac{-(3+\alpha)}{9+\alpha}}\frac{\phi}{M_{pl}} \Big) \bigg]
              ^{ - \frac{1}{3}\frac{9+\alpha}{-(3+\alpha)}}} {
        \sqrt{ -(6+\alpha) +
           \alpha \cos \Big( 2\sqrt{6} \sqrt{\frac{-(3+\alpha)}{9+\alpha}}\frac{\phi}{M_{pl}}\Big)
            +  2(3+\alpha)  \sin \Big( \sqrt{6} \sqrt{\frac{-(3+\alpha)}{9+\alpha}}\frac{\phi}{M_{pl}}\Big)
              } } \ ,
\end{eqnarray}
where we assumed $-9 < \alpha < -3$ .
Taking a look at Eq.(\ref{eq12}), we see that there are apparently two solutions for a given Hubble parameter $H_{a}$. 
However, since the two branches stemming from Eq.(\ref{eq12}) are related by   
the field redefinition of the scalar field, there is only one solution.
Taking the parameter close to that for the slow roll inflation, $\alpha \simeq -3$, 
we can realize slow roll inflation with anisotropic expansion in principle.
The anisotropic expansion becomes a prolate type because $n$ is negative. 
However, for inflation to occur, we have to require $V(\phi)$ to be positive, 
which makes the gauge field a ghost, namely $f(\phi)$ becomes imaginary. Interestingly, 
in this case, we can numerically confirm that this anisotropic constant-roll solution can not be an attractor.

Next, we consider another example $C_{1}=M/2, C_{2}=- M/2$, namely, 
\begin{equation}
  H_{a}(\phi) = M \sinh \Bigg( \sqrt{6} \sqrt{\frac{3+\alpha}{9+\alpha}} \frac{\phi}{M_{pl}} \Bigg) \ .
    \label{eq21}
\end{equation}
The solution (\ref{eq21}) gives rise to 
\begin{eqnarray}
  \dot{\phi} = M M_{pl} \sqrt{\frac{(3+\alpha)(9+\alpha)}{6}} 
       \bigg[ \pm 1 - \cosh \Big( \sqrt{6} \sqrt{\frac{3+\alpha}{9+\alpha}}\frac{\phi}{M_{pl}} \Big) \bigg]
\end{eqnarray}
and
\begin{eqnarray} 
      a  \propto  \bigg[ 
           \mp 1 + 
           \cosh \Big( \sqrt{6} \sqrt{\frac{3+\alpha}{9+\alpha}}\frac{\phi}{M_{pl}} \Big) 
              \bigg]^{-\frac{1}{3+\alpha}} \quad .   \label{eq22}
\end{eqnarray}
Thus, we can determine the following functional forms:
\begin{eqnarray}
  V &=& \frac{M^{2} M_{pl}^{2}(\alpha-9)}{24}  \bigg[  (6+\alpha) +
             \alpha \cosh \Big( 2\sqrt{6} \sqrt{\frac{3+\alpha}{9+\alpha}}\frac{\phi}{M_{pl}}\Big) \nonumber \\ 
             && \hspace{4cm} \mp
              2(3+\alpha) \cosh \Big( \sqrt{6} \sqrt{\frac{3+\alpha}{9+\alpha}}\frac{\phi}{M_{pl}} \Big) 
              \bigg] \ ,  \label{eq299}
\end{eqnarray}
and 
\begin{eqnarray}
 f \propto \frac{ \bigg[   
            \mp 1 + 
           \cosh \Big( \sqrt{6} \sqrt{\frac{3+\alpha}{9+\alpha}}\frac{\phi}{M_{pl}} \Big) 
              \bigg]^{\frac{1}{3}\frac{9+\alpha}{3+\alpha}}}{
           \sqrt{ - (6+\alpha) -
           \alpha \cosh \Big(  2\sqrt{6} \sqrt{\frac{3+\alpha}{9+\alpha}}\frac{\phi}{M_{pl}} \Big)
           \pm 2(3+\alpha) \cosh \Big(  \sqrt{6} \sqrt{\frac{3+\alpha}{9+\alpha}}\frac{\phi}{M_{pl}} \Big)
           } }   \ ,
\end{eqnarray}
where we assumed $-3 < \alpha$.
Again, inflation can occur taking the parameter close to that of the slow roll inflation $\alpha \simeq -3$.
In this case, anisotropic expansion has an oblate type because $n$ is positive. 
It turns out that these are viable models.

\begin{figure}[h]
\includegraphics[width=7cm,angle=270]{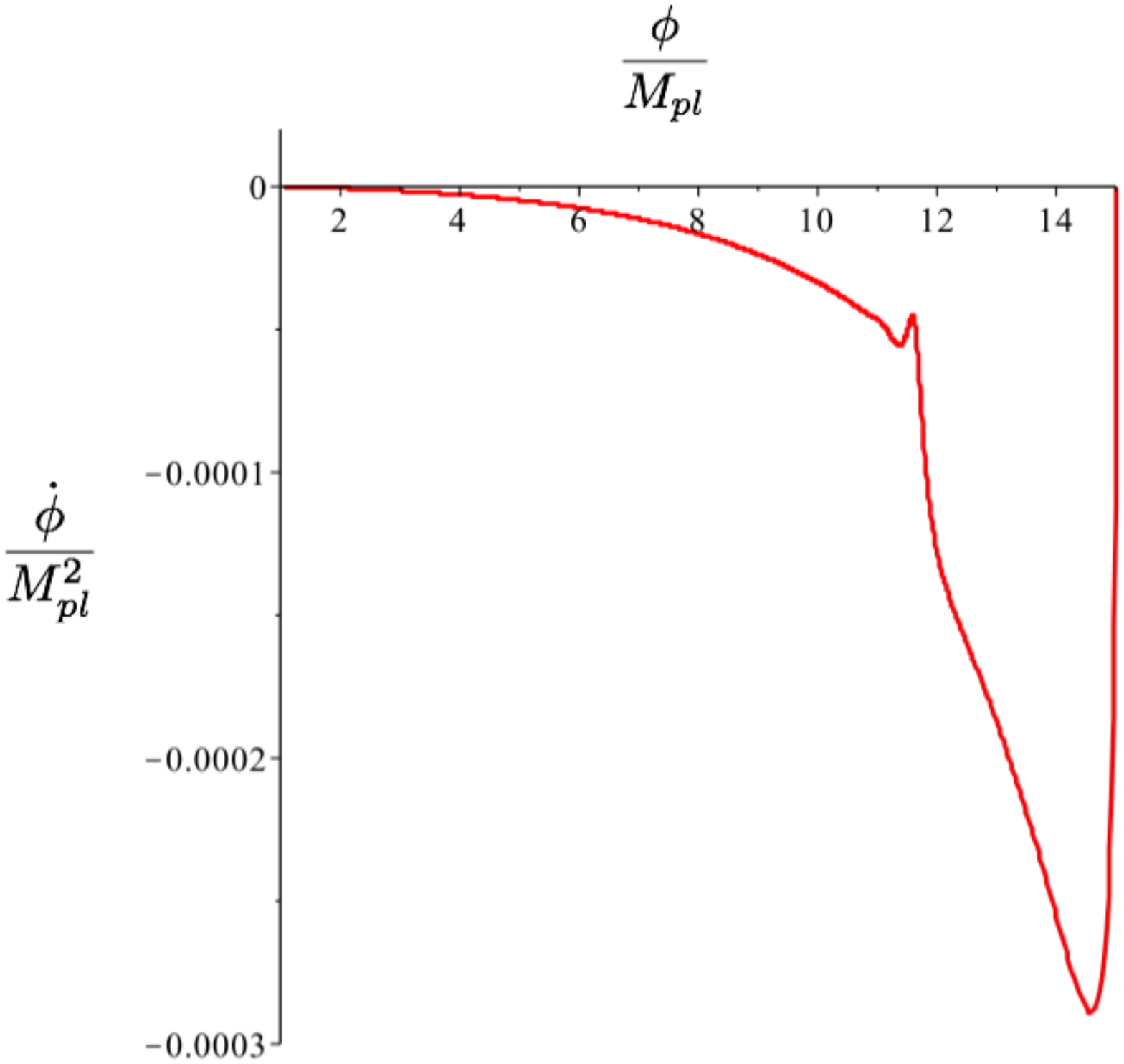}
\caption{The dynamics of the inflaton field is depicted in the phase space.
The initial conditions are set 
$\frac{\phi_{0}}{M_{pl}} = 15  , \frac{\dot{\phi}_{0}}{M_{pl}^{2}} = 0$.
Around $\frac{\phi}{M_{pl}} = 12$, the back reaction from the gauge field
starts to affect the inflaton and eventually it converges to the solution.
}
\label{fig1}
\end{figure}

\section{\normalsize{P\lowercase{hase space dynamics}}}

\begin{figure}[h]
\includegraphics[width=7cm,angle=270]{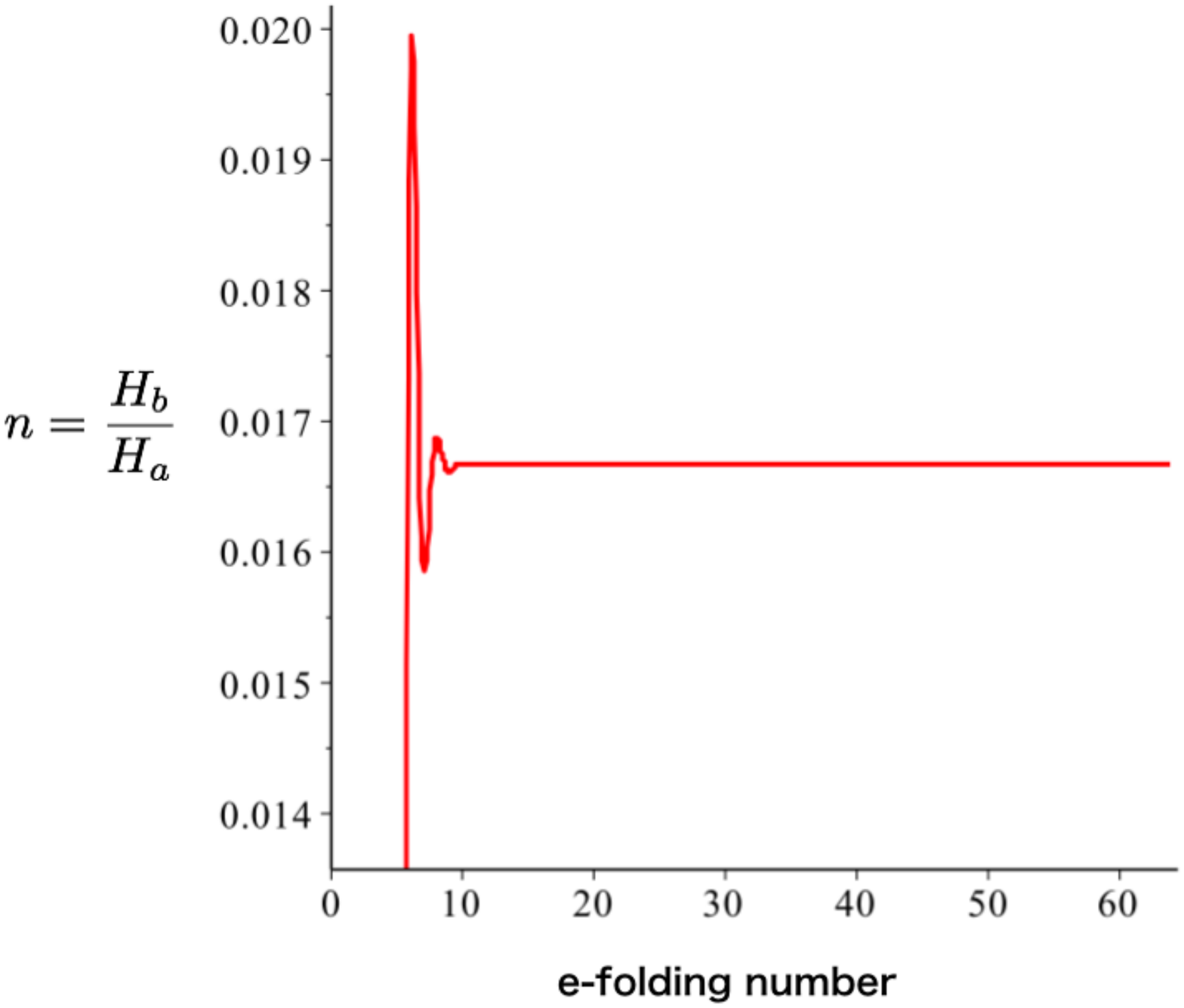}
\caption{The evolution of the degree of the anisotropic expansion
$ n = \frac{H_{b}}{H_{a}} $ to the e-folding number is depicted.
Around $ \frac{\phi}{M_{pl}} = 10 $, it converges to the solution $ \frac{3+\alpha}{6} \simeq 0.0167 $ .
}
\label{fig2}
\end{figure}

\begin{figure}[h]
\includegraphics[width=7cm,angle=270]{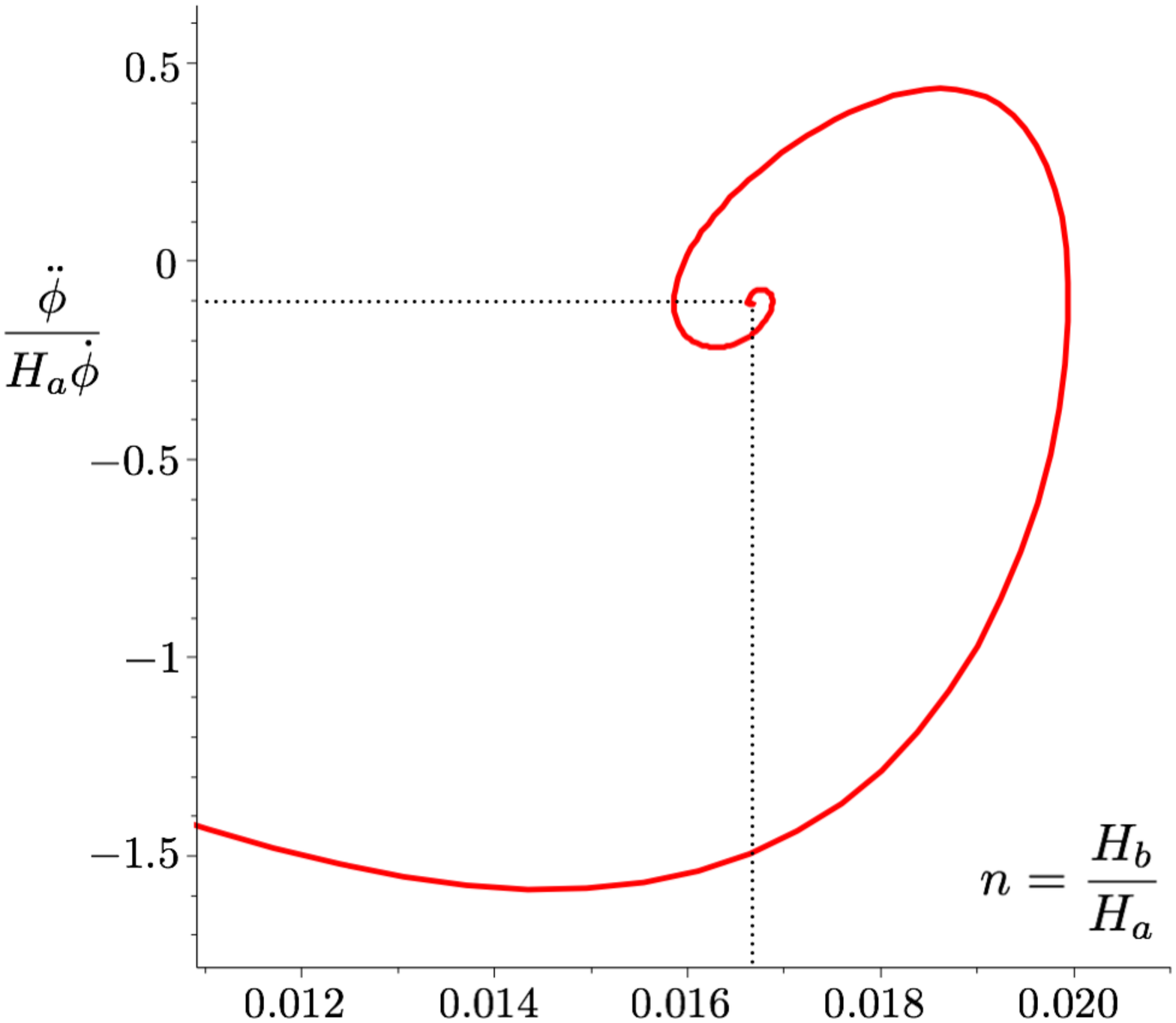}
\caption{The trajectory in  $\frac{\ddot{\phi}}{H_{a}\dot{\phi}}$ and $ n = \frac{H_{b}}{H_{a}} $ plane
is shown.
One can clearly see the anisotropic constant-roll inflation is an attractor.}
\label{fig3}
\end{figure}

To see if the solutions derived from Eq.(\ref{eq21}) are attractors, let us solve 
Eqs.(\ref{eq4})-(\ref{klein}) numerically.  
Using the upper sign in Eq.(\ref{eq299}) as the potential $V(\phi)$ 
and the coupling function $f(\phi)$, 
we obtained Figs.\ref{fig1},\ref{fig2} and \ref{fig3}.
In these numerical calculations, we have chosen parameters $\alpha = -2.9 \ , \ M = 10^{-5} M_{pl}$.

In Fig.\ref{fig1}, we plotted a trajectory in the phase space of the inflaton.
From Fig.\ref{fig2}, we can see the anisotropy converges to a constant value $n$. 
In Fig.\ref{fig3}, 
we see that the rate $\frac{\ddot{\phi}}{H_{a}\dot{\phi}}$ and the anisotropy $n=\frac{H_{b}}{H_{a}}$
become constant after e-folding number has reached about $10$ in the phase space.
This is certainly the anisotropic constant-roll inflationary solution.
Indeed, the trajectory is approaching exactly predicted values $\frac{\ddot{\phi}}{H_{a}\dot{\phi}} = -(3+\alpha) = -0.1$ and 
$n = \frac{3+\alpha}{6} \simeq 0.0167$.
Thus, from Figs.\ref{fig1}-\ref{fig3}, it turned out that the solution is an attractor.
We also confirmed that the behavior is independent of initial conditions.
Note that the gauge field suffers from a strong coupling problem ($f(\phi) < 1$) around $\phi = 0$. 
Thus, inflation needs to finish before $\phi$ approaches the origin.

On the other hand, in the case that we choose the lower sign in Eq.(\ref{eq299}), 
there is no strong coupling problem.
Instead, when $\alpha \simeq -3$, the potential is negative around $\phi = 0$ and then 
inflation needs to finish before $\phi$ approaches the origin even for this case. 
Otherwise, the solution is useful.
Indeed, we confirmed that this anisotropic constant-roll inflationary solution is also an attractor. 
\section{\normalsize{S\lowercase{ummary}}}
We studied  constant-roll inflation  in the presence of a gauge field coupled to an inflaton. 
We imposed the constant-roll condition, $\frac{\ddot{\phi}}{H \dot{\phi}} = $ const., and
the constant anisotropy condition.
Then we looked for anisotropic inflationary solutions.
Remarkably, we found exact solutions which describe anisotropic constant-roll inflation.
Our result includes anisotropic power-law inflation~\cite{Kanno:2010nr},\cite{Ito:2015sxj} as a special case.
We also numerically demonstrated that
our new solutions are attractors in the phase space.

It should be noted that in our model there is a relation, Eq.(\ref{consistency}), between 
the slow roll parameters and the anisotropy $n$.
$n$ must be very small on the CMB scales 
and then the slow roll parameters also become too small. 
As a result, it is difficult to construct all periods of inflation using our model
if the inflaton is responsible for the curvature perturbations.
Our model could be viable on small scales (even if on the CMB scales, $l \sim 1000$).
In such a case, our model predicts interesting observables such as the statistical anisotropy, which could 
take a larger value compared with that on the large scales observationally.

Although we have considered anisotropic inflation with a gauge field, 
our method to find exact solutions can easily be extended to the 
anisotropic inflationary models with a two-form field \cite{Ito:2015sxj}-\cite{Ohashi:2013qba}.
In the case that a gauge field and a two-form field coexist,
it is known that there appears an exact solution where both fields survive and an
isotropic universe is realized for a particular parameter set~\cite{Ito:2015sxj}.
This is because a gauge field and a two-form field produce opposite anisotropy.
It is also interesting to apply our method to anisotropic inflation with multi-gauge field (two-form field) models
\cite{Yamamoto:2012tq},\cite{Yamamoto:2012sq}.
We leave these issues for future work.

\begin{acknowledgments}
A.I. would like to thank Hayato Motohashi for helpful discussions.
A.I. was supported by Grant-in-Aid for JSPS Research Fellow and JSPS KAKENHI Grant No.JP17J00216. 
J.S. was in part supported by JSPS KAKENHI Grant Numbers JP17H02894, JP17K18778, JP15H05895, JP17H06359.
\end{acknowledgments}

\end{document}